\documentclass[12pt,preprint]{aastex}

\slugcomment{PASP, accepted}

\shorttitle{Spectrophotometry with a transmission grating}
\shortauthors{Kenworthy and Hinz}

\begin{document}

\title{Spectrophotometry with a transmission grating \\
   for detecting faint occultations}

\author{M.A. Kenworthy}
\affil{Physics Department, University of Cincinnati, Cincinnati, OH
45221}
\email{matt@physics.uc.edu}

\and

\author{P.M. Hinz}
\affil{Steward Observatory, University of Arizona, 933 N. Cherry Avenue,
Tucson, AZ 85721}
\email{phinz@as.arizona.edu}

\begin{abstract}

High-precision spectrophotometry is highly desirable in detecting and
characterizing close-in extrasolar planets to learn about their makeup
and temperature.  For such a goal, a modest-size telescope with a simple
low-resolution spectroscopic instrument is potentially as good or better
than a complex general purpose spectrograph since calibration and
removal of systematic errors is expected to dominate. We use a
transmission grating placed in front of an imaging CCD camera on Steward
Observatory's Kuiper 1.5 m telescope to provide a high signal-to-noise,
low dispersion visible spectrum of the star HD 209458. We attempt to
detect the reflected light signal from the extra-solar planet HD 209458b
by differencing the signal just before and after secondary occultation.
We present a simple data reduction method and explore the limits of
ground based low-dispersion spectrophotometry with a diffraction
grating.  Reflected light detection levels of 0.1\% are achievable for
$5000-7000$~\AA, too coarse for useful limits on ESPs but potentially
useful for determining spectra of short-period binary systems with large
$(\Delta m_{vis}=6)$ brightness ratios. Limits on the precison are set
by variations in atmospheric seeing in the low-resolution spectrum.
Calibration of this effect can be carried out by measurement of
atmospheric parameters from the observations themselves, which may allow
the precision to be limited by the noise due to photon statistics and
atmospheric scintillation effects.

\end{abstract}


\section{Introduction}

The indirect detection of large hot Jupiter-like extrasolar planets with
periods on the order of days raises the possibility of direct detection
of the reflected light from the planet's atmosphere. The expected
contrast ratio for these close-in gas giant planets is expected to be
quite favorable $(10^4-10^5)$ compared to the contrast ratio between
Jupiter and our sun $(10^{9})$.  However, the close proximity of these
planets to their stars precludes spatially resolving the planet with
current telescopes. Thus additional information about these planets
requires other methods of discerning its flux in the presence of the
much brighter star. Several approaches are possible to separate the
starlight from the planet's light.  \cite{cha99} and \cite{col01} each
observed tau Boo with a high resolution spectrometer in an attempt to
discern the planet's light through by the Doppler shift of the reflected
light as the planet orbits its star.  Although no detection was
observed, the data were sufficient to place upper limits on the planet's
albedo.

The discovery that the planet around HD 209458 occulted its star during
the 3.5 day orbit \citep{cha00} was a significant step in verifying the
existence of the extrasolar planet candidates. \cite{bro01} followed up
this initial detection with high precision HST photometric measurements
of the transit. This yielded a more precise light curve of the
occultation, and gave the first indication of the composition of the
planet's atmosphere. The transit was detected to be slightly deeper at
the wavelength of sodium absorption \citep{cha02}, as predicted by
various groups \citep{bro01a,sea00,hub01}.

Direct detection of the reflected light from a close-in planet has the
potential of telling us the albedo of a planet, and thus also, the
temperature of the planet.  From a more pedestrian viewpoint detection
of the planet's reflected light would be the indication of what the
planet ``looks'' like. The color is more than just a curiosity, though,
allowing the first constraints on the planet's chemical makeup and
vertical structure (such as the existence of clouds).

We observe the HD209458 system just before and after the eclipse of the
planet by its parent star.  Theoretical models suggest the planet may
have a bright (0.7) albedo with a wide ($R/\Delta R\sim 20$) absorption
feature due to pressure broadened sodium at 591 nm \citep{sud00}.  The
reflected light fraction from the star is $2\times10^{-4}$ for a
geometric albedo of one, so a detection level of $\simeq 1\times10^{-4}$
per wavelength band is needed to produce a useful constraint on the
albedo of the planet.

Colour information is obtained with the use of a transmission grating in
front of a CCD detector, in an objective prism configuration. Objective
prism surveys are used with wide field imaging telescopes for spectral
typing of stars and classifications of the brighter galaxies. The low
dispersion, high throughput and no slit jaw losses make the objective
prism method suitable for colour dependent transits. The main challenge
is then in obtaining accurate spectrophotometry with the variation in
transmission of the atmosphere as a function of airmass and the
variation of spectral resolution due to the local seeing at the
telescope.

\section{Observations}

All our observations were taken at the Kuiper telescope on Mount
Bigelow, North of Tucson, a 1.54 m reflecting telescope with a
CCD camera mounted at the Cassegrain focus.

The grating is a 72 lines/mm epoxy layer stamped onto glass $24\times 24
\times 3$mm, manufactured by Edmunds Scientific. The ruling is blazed to
give maximum transmission ($32\%$) in the first order. The grating is
placed in the camera filter wheel approximately 10 cm from the detector.

The CCD is an unthinned $2048^2$ pixel detector manufactured by the
University of Arizona Imaging Technology Laboratory. The camera scale is
0.15 arcseconds per pixel and the unbinned spectral dispersion of the
transmission grating at 5000\AA~is approximately 20\AA/pixel.  We
on-chip bin the CCD to 3 by 3 pixels, to maximize the readout rate of
the detector, while still retaining the full spatial and spectral
resolution for the typically 1 arcsec images at the Kuiper telescope.
The distance between the grating and detector results in a spectral
resolution of 39 at 5000A in 1 arcsecond seeing.  Exposures were set at
5 seconds to give a peak flux integration of 50,000 ADUs, to keep the
detector in a linear response range and to avoid going out of range on
the 16 bit analog-to-digital converter for the CCD.

\begin{figure}
\includegraphics[angle=270,width=\columnwidth]{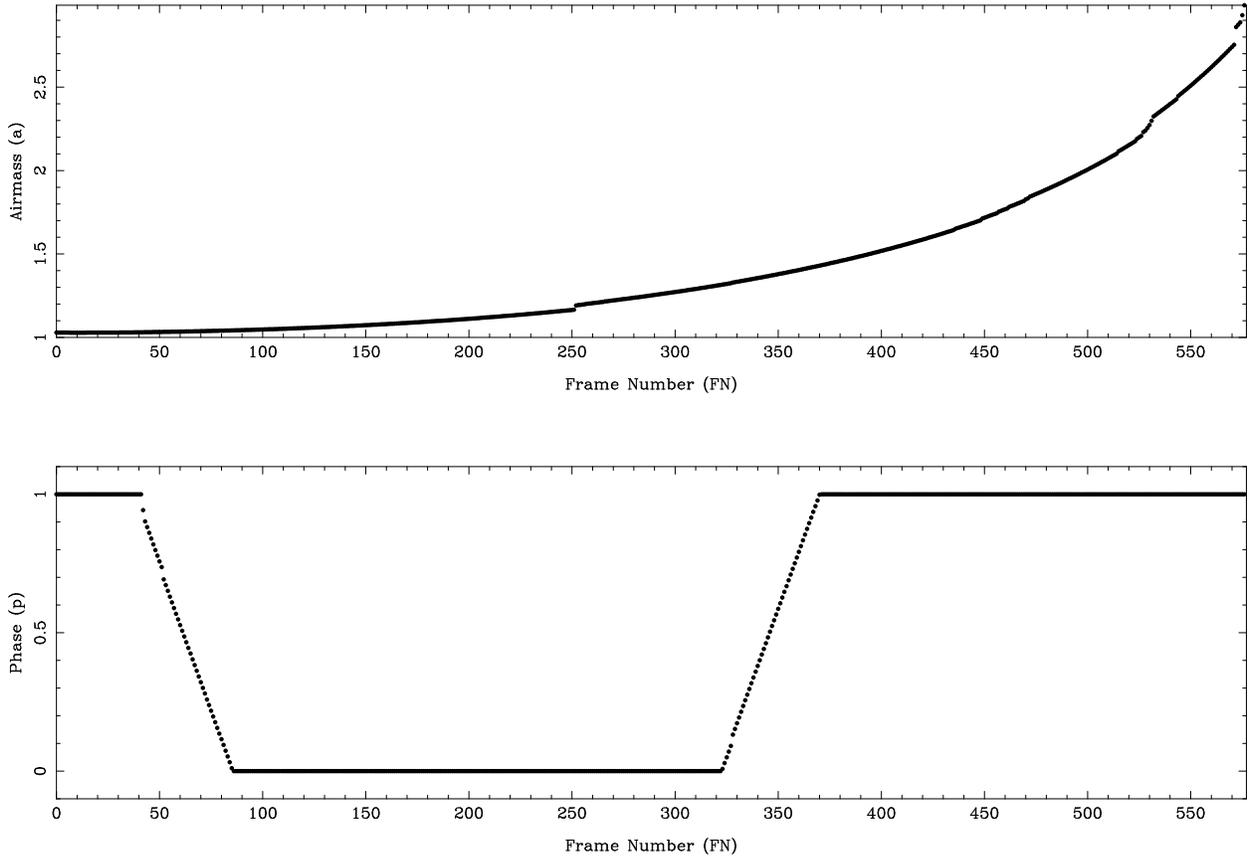}

\caption{The variation of airmass and planet phase with Frame
Number\label{airmass}}

\end{figure}

The secondary eclipse of the planet lasts three hours, with a period of
just over 3.5 days \citep{bro01}.  The data presented are from 12
November 2001 UT, which was a calm photometric night during which the
secondary eclipse started soon after the star crossed the meridian (see
Table \ref{ephem_table}).  Over 600, five second frames of HD 209458
were acquired as the star began setting, going from airmass 1.03 to 3.0
during the observations (see Figure \ref{airmass}). Approximately equal
numbers of frames were obtained of the star in and out of secondary
eclipse. For all observations the telescope was manually guided to keep
the spectrum in approximately the same part of the CCD. This number was
reduced to 575 after rejection of frames where the flux of Star 1
(discussed later) showed a significant drop, most likely due to thin
cloud passing overhead late in the evening.

\section{Throughput}

\begin{figure}
\includegraphics[angle=270,width=\columnwidth]{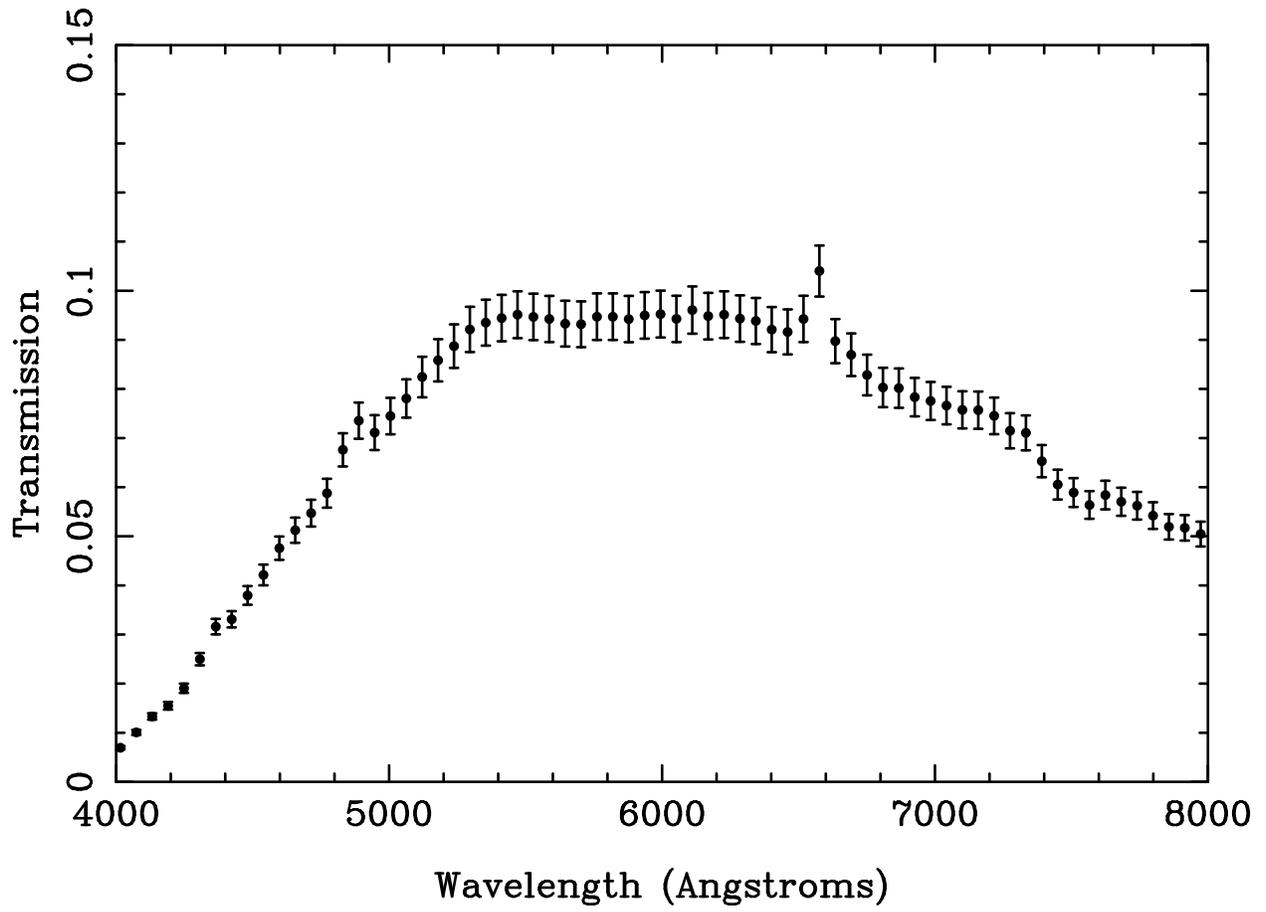}
\caption{The measured efficiency for the instrument as a function of
wavelength, corrected to an airmass of 1.\label{thput}}
\end{figure}

An observation of a WD primary spectrophotometric standard GD 71
\citep{boh95} provides an estimation of the throughput of the
instrument. The spectrum of the standard star is corrected to an airmass
of one, and Figure \ref{thput} gives the estimated total throughput from
above the atmosphere through to counts in the detector. The estimated
contributions to the throughput are presented in Table
\ref{thput_table}. The dominant errors for the calculation are from the
uncertainty in the gain factor for the CCD.

\section{Colour response of the pixels}

The gain ,$g$, of modern CCDs is assumed to be a function of pixel
position $x,y$ and of wavelength $\lambda$, making $g=g(x,y,\lambda)$.
In imaging the assumption is that the gain is achromatic
($g(x,y,\lambda)=g(x,y)$) for a given fixed passband.  A spectrograph
uses a dispersing element to convert wavelength into spatial position on
the detector, i.e.  $g=g(x,y,\lambda(x,y))$ where $\lambda(x,y)$ is a
well defined function fixed by the position of the dispersing element
and the entrance slit.

Objective grating spectroscopy lies between these two extremes. The
``entrance slit'' is defined by the position and image size of the
object in the sky, and any image motion of the object results in a
motion of its spectrum on the CCD. This means that $\lambda(x,y)$ is not
a well-defined function. Without a slit in the system we need to apply a
gain correction which is based on white light illumination and assume
this is valid for the range of wavelengths observed with the grating. 

To see if our assumption of an achromatic gain ($g(x,y,\lambda)=g(x,y)$)
is valid for our CCD we took two sets of flat fields with different
colours, one with the telescope dome open and the telescope pointing to
the morning twilight sky (a predominantly blue colour) and the other
with the dome closed and tungsten lights switched on (a 3000 K
blackbody) and pointing at the dome ceiling. Both sets were taken with
the filter wheel rotated to the ``clear'' slot.

For the twilight (sky) flats, frames with mean counts above 50000 or
lower than 10000 were rejected, along with any that have an exposure
time shorter than 2 seconds, to prevent illumination gradients due to
the finite shutter speed.  For the tungsten lamp (dome) flats, a set of
114 exposures were taken with a mean count level of 50,000.  For both
sky and dome flats, the individual frames are scaled individually by
their respective modes and then combined as an average with avsigclip
rejection. 

\begin{figure}
\includegraphics[angle=270,width=\columnwidth]{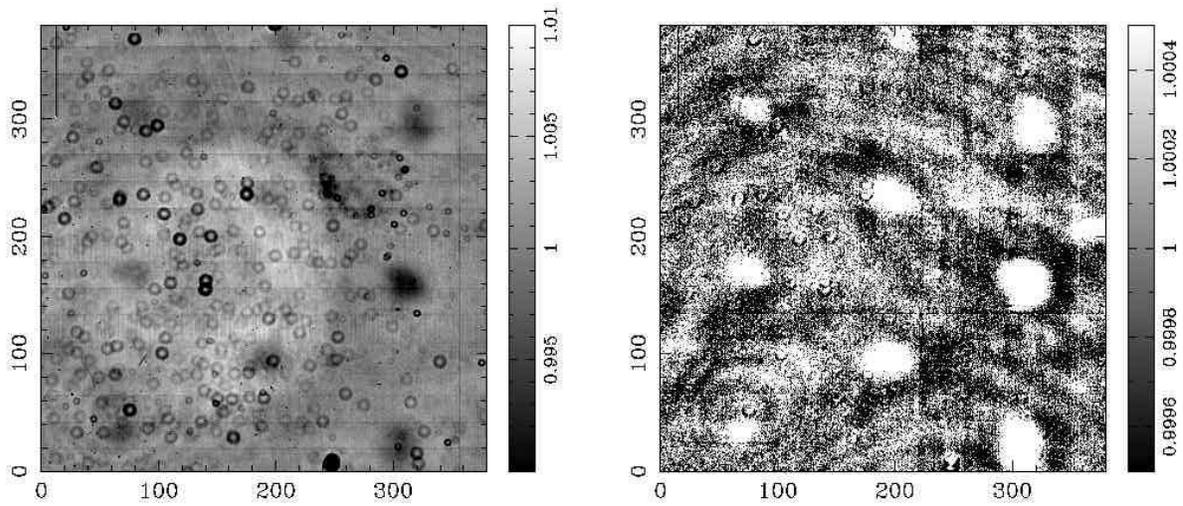}

\caption{The Dome flat field is shown in the left-hand image, and the
dome flat divided by the sky flat is shown in the right-hand image. The
region of the CCD that images the spectrum shows no significant
variations between sky and dome illumination above the noise in the two
frames, indicating that the gain for the detector is achromatic and
either frame can be used for gain correction of the
spectrum.\label{flats}}

\end{figure}

The two flat fields are processed by \tt MKSKYFLAT \rm to remove large
scale illumination effects. Both flats show CCD manufacturing
imperfections and 'mirror donuts' symptomatic of dust on the dewar
window surfaces, and dividing the dome flat by the sky flat removes
virtually all the larger dust artefacts (see left-hand of Figure
\ref{flats}).  Given the quoted gain of $3.2\pm0.1 e^-/ \rm{ADU} $ for
the CCD the statistics shown in table \ref{flats_table} are consistent
with photon shot noise being the dominant source of noise in the
combined flats.

Dividing the dome flat into the sky flat produces the image shown in the
right-hand side of Figure \ref{flats}. The residual artefacts are a
triangularly spaced set of higher-gain spots, and some fringing that is
just visible in the noise of the image. The spots are thought to be
where adhesive holds the CCD in its protective housing, and the fringes
are variations in the doping applied in the manufacturing process of the
CCD. 

Measuring the fractional noise in flat regions of this image gives a
value of $9.0\pm1.0\times10^{-4}$, consistent with the noise level of
$8.17\pm1.56\times10^{-4}$ expected when dividing the sky flat into the
dome flat. This image shows that, apart from some obvious regions of the
CCD which can be avoided, both the sky flat and dome flat agree with
each other within errors. The lack of any major gradient across the CCD
means that $g(x,y,\lambda)=g(x,y)$ is valid to within measurable errors
and the flat with the higher S/N ratio can be used without introducing
any colour biased systematic noise into the data. We can therefore use a
``white light'' flat to correct for gain variations over the whole
spectrum.

\section{Data Reduction}

\subsection{Frame reduction}

The data were reduced with the \tt IRAF \rm data reduction tasks. 100
zero frames are combined with the CCDPROC task using the ccdclip
rejection for removing outlying events, and this combined zero frame is
subtracted from the data. The data is then trimmed to remove the bias
and overscan regions of the CCD.

\begin{figure}
\includegraphics[angle=270,width=\columnwidth]{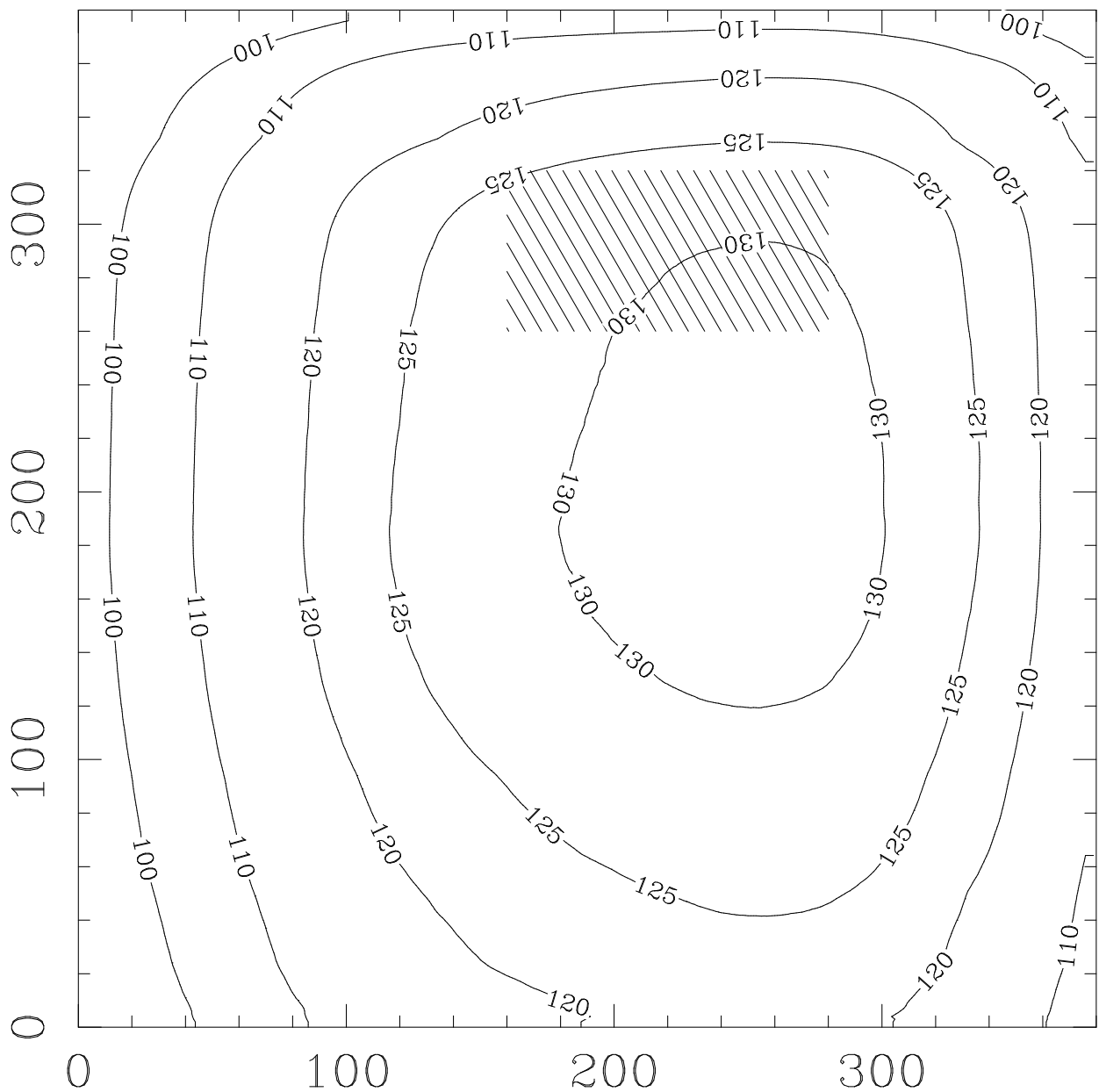}

\caption{ Contour plot of the sky background flux. The sky background is
not uniform due to the vignetting of the grating. The contours show
light levels after the image is median smoothed. The hatched region
indicates the location of the spectra. North is up and East to the
left.\label{scattered}}

\end{figure}

\begin{figure}
\includegraphics[angle=0,width=\columnwidth]{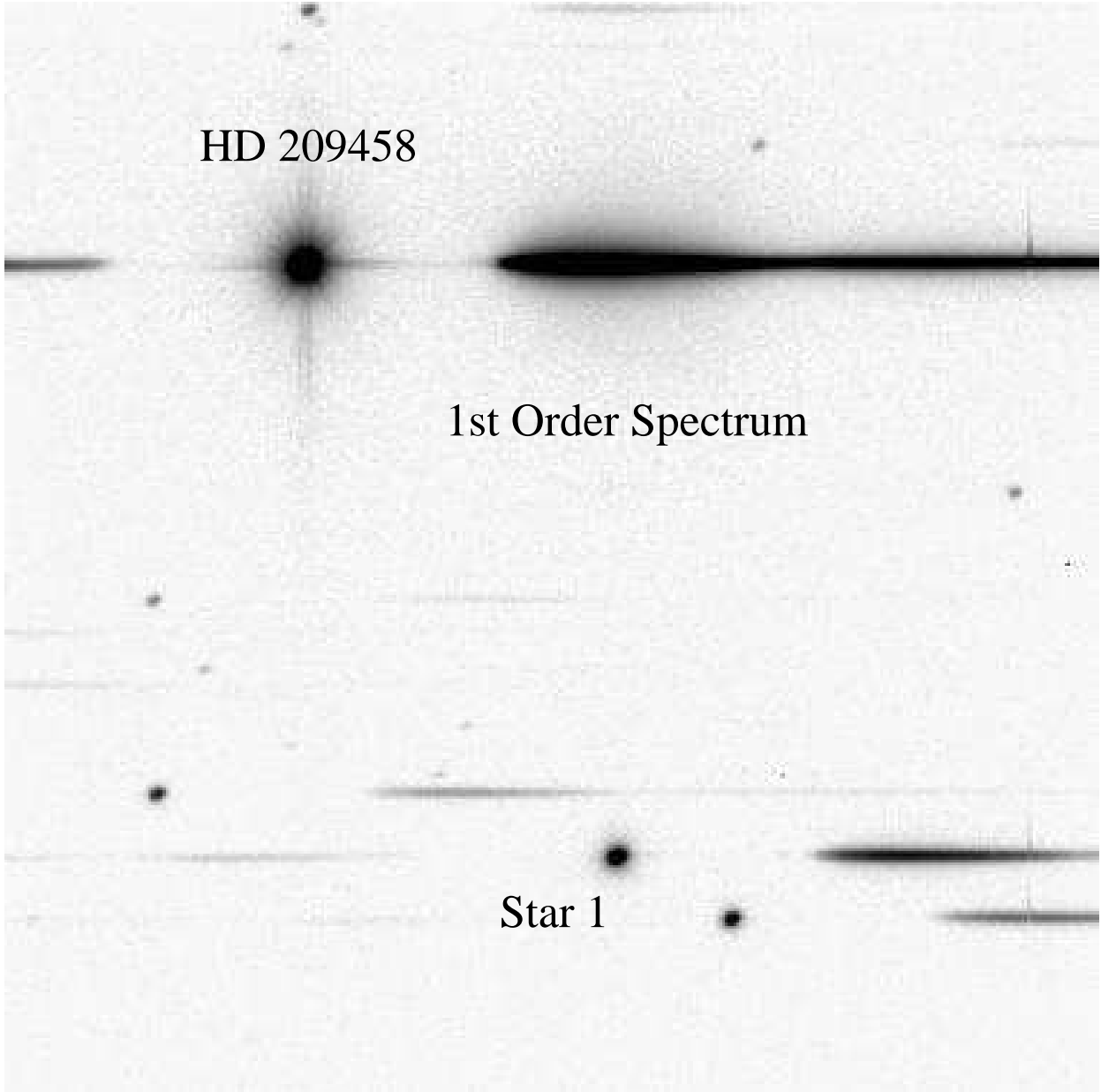}
\caption{A typical data frame showing the positions of HD~209458 and the
star used in estimating the scintillation noise. North is up and East to
the left. \label{fframe}} \end{figure}

The data frames are flat-fielded using a gain map constructed from the
dome flat field described in the previous section. The grating creates a
non-uniform illumination pattern across the detector, due to vignetting
of the incoming light.  To remove this a sky background frame is
generated by combining 100 frames taken with the telescope drive
switched off. These frames are median scaled and median combined with
the ccdclip rejection algorithm. The resulting image is filtered with a
boxcar median filter 51 pixels to a side to smooth out high spatial
frequency noise (see Figure \ref{scattered}).  Differencing this
smoothed image from the unsmoothed original shows no remaining large
scale variations in the sky background image.  This sky background frame
is then scaled to a blank region of the CCD and subtracted on a per
frame basis.  Bad columns and hot pixels are replaced by interpolation
from the nearest neighbouring pixel values with \tt FIXPIX \rm (see
Figure \ref{fframe}). 

The plate scale is determined from the position of three stars in the field
of view from an image taken at an airmass of 1.03. Centroid fitting for
each of the stars is performed, and the relative separations of the
stars are taken from the astrometric solution attached to the DSS image
\footnote{The Digitized Sky Survey was produced at the Space Telescope
Science Institute under U.S. Government grant NAG W-2166. The images of
these surveys are based on photographic data obtained using the Oschin
Schmidt Telescope on Palomar Mountain and the UK Schmidt Telescope. The
plates were processed into the present compressed digital form with the
permission of these institutions.} covering the region around
HD~209458.  The plate scale is calculated to be $0.4314\pm0.0015
\arcsec/\rm{pix}$.

We use the three brightest stars in the frame to provide registration
between the frames. The \tt IRAF \rm routine \tt IMCENTROID \rm gives
the frame offset relative to the first frame of the night. 

The total range of image motion is (8.5,5.8) pixels in the (RA,dec)
axes, with typical (RMS) shift of $\delta RA_{rms}=1.119, \delta
dec_{rms} = 0.823$ pixels, less than the typical seeing disk size of 3
pixels (1 arcsecond). The larger drift in the RA axis is thought to be
due to periodic corrections in the siderial rate applied by the
telescope guidance computer when a certain drift limit is exceeded.

\begin{figure}
\includegraphics[angle=270,width=\columnwidth]{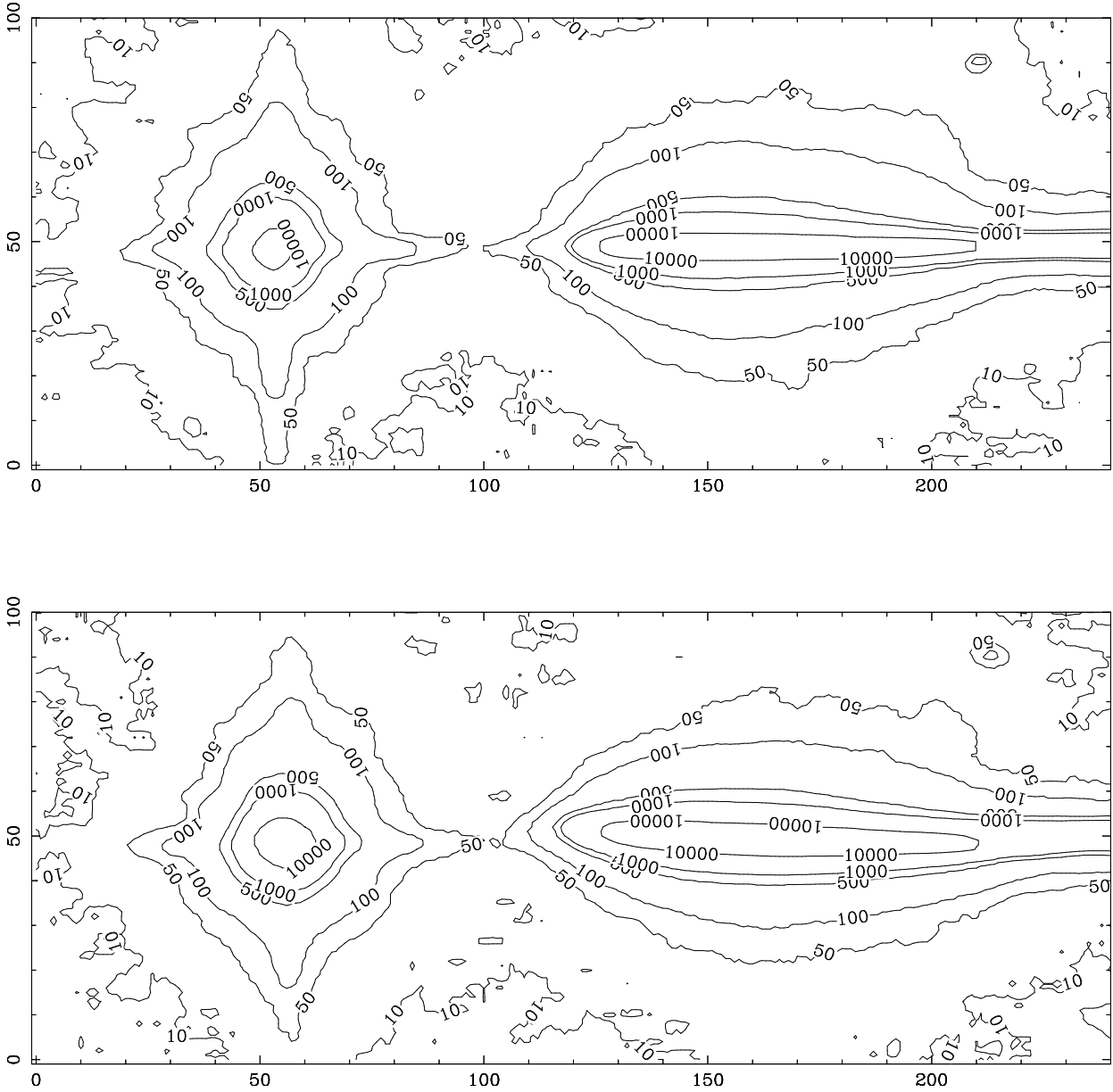}

\caption{Scattered light from the grating. The zeroth order image and
first order spectrum are seen. These images have been smoothed with a
Gaussian and are in units of photons. The upper image is from airmass
1.03, the lower image taken at an airmass of 2.7. Both have had the sky
background and detector bias levels subtracted. North is up and East to
the left.\label{grat_scattered}}

\end{figure}

The remaining sources of light are the zeroth order images, the first
and second order spectra from the grating, and a halo of scattered light
surrounding the spectra themselves.  A typical sky background subtracted
image is shown in Figure \ref{grat_scattered}, with a contour plot
showing the scattered light profile around the spectrum. As can be seen
in Figure \ref{grat_scattered}, the scattered light around the spectrum
is constant for images at the beginning and end of the night, indicating
it does not effect our measurement of the light in the core of the
spectrum in a systematic way.

\subsection{Spectrum extraction and calibration}

\begin{figure}
\includegraphics[angle=270,width=\columnwidth]{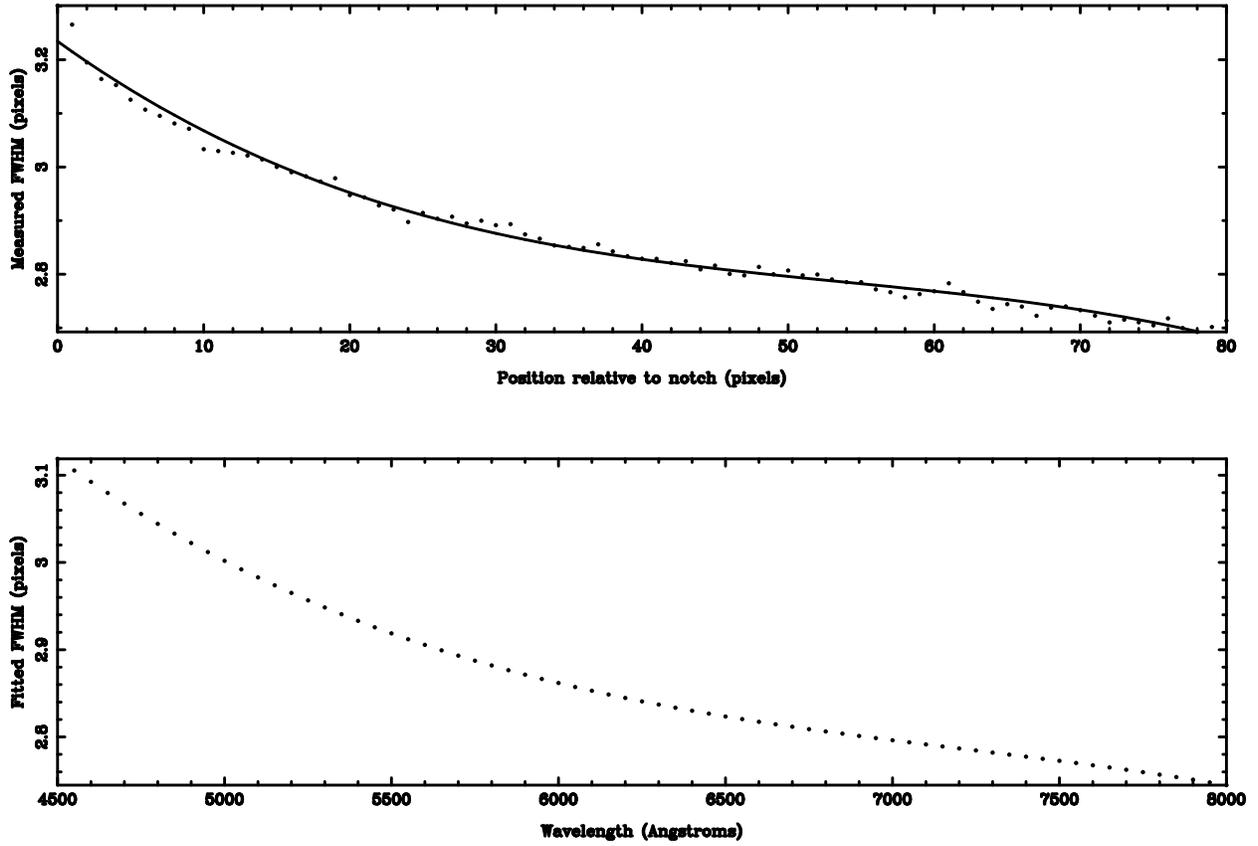}

\caption{The measured FWHM as a function of pixel position along the
spectum (upper figure) and the fitted FWHM position as a function of
wavelength after dispersion correction (lower figure) for Frame Number
1.  \label{fwhm}}

\end{figure}

\begin{figure}
\includegraphics[angle=270,width=\columnwidth]{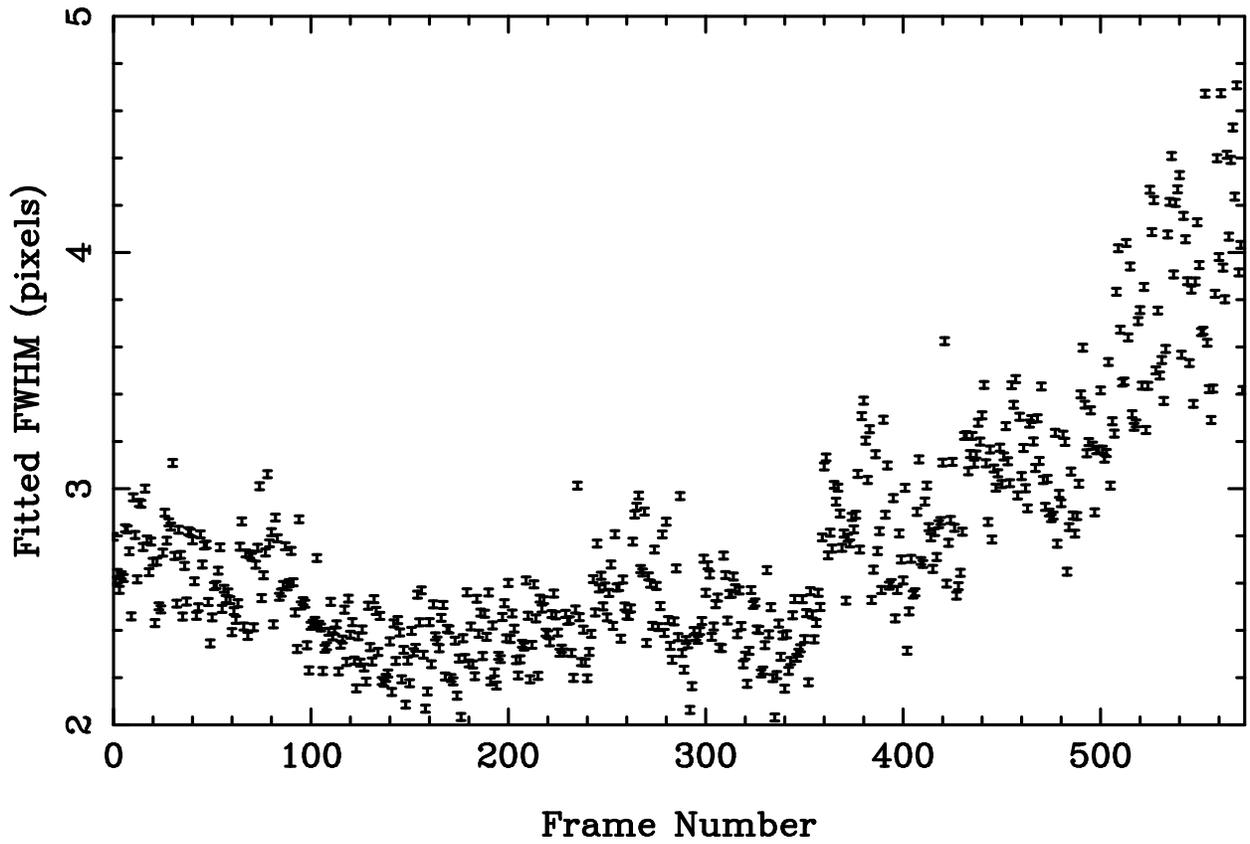}

\caption{The measured transverse FWHM at 7000\AA. For clarity, the
points themselves are not shown, only their error bars.
\label{fittedfwhm}}

\end{figure}

Using the centroid of Star 1 as a fiducial, a box is drawn around and
centered on the first order spectrum, where the columns are
perpendicular to the dispersion axis of the grating. A Gaussian curve is
fit along each column to measure the PSF of the grating at that
wavelength and to trace the centroid of the spectrum across the CCD. A
low-order polynomial is fit to both the measured FWHM and centroid
positions (see Figure \ref{fwhm}). The RMS between the FWHM fit and data
is $1.48\times 10^{-2}$ pixels for all frames during the evening. A plot
of the transverse FWHM at 7000\AA~ as a function of Frame Number (Figure
\ref{fittedfwhm}) shows that the seeing varies on timescales of seconds,
minutes and hours during the observations.

The counts within each column are summed to give a one-dimensional
spectrum with flux per wavelength bin. The width of the box is set at 41
pixels (17 arcsec), extending well into the scattered light region from
the grating.  Increasing the extraction width of the box does not change
the extracted planetary spectrum, implying that our assumption about the
scattered light is valid.

\begin{figure}
\includegraphics[angle=270,width=\columnwidth]{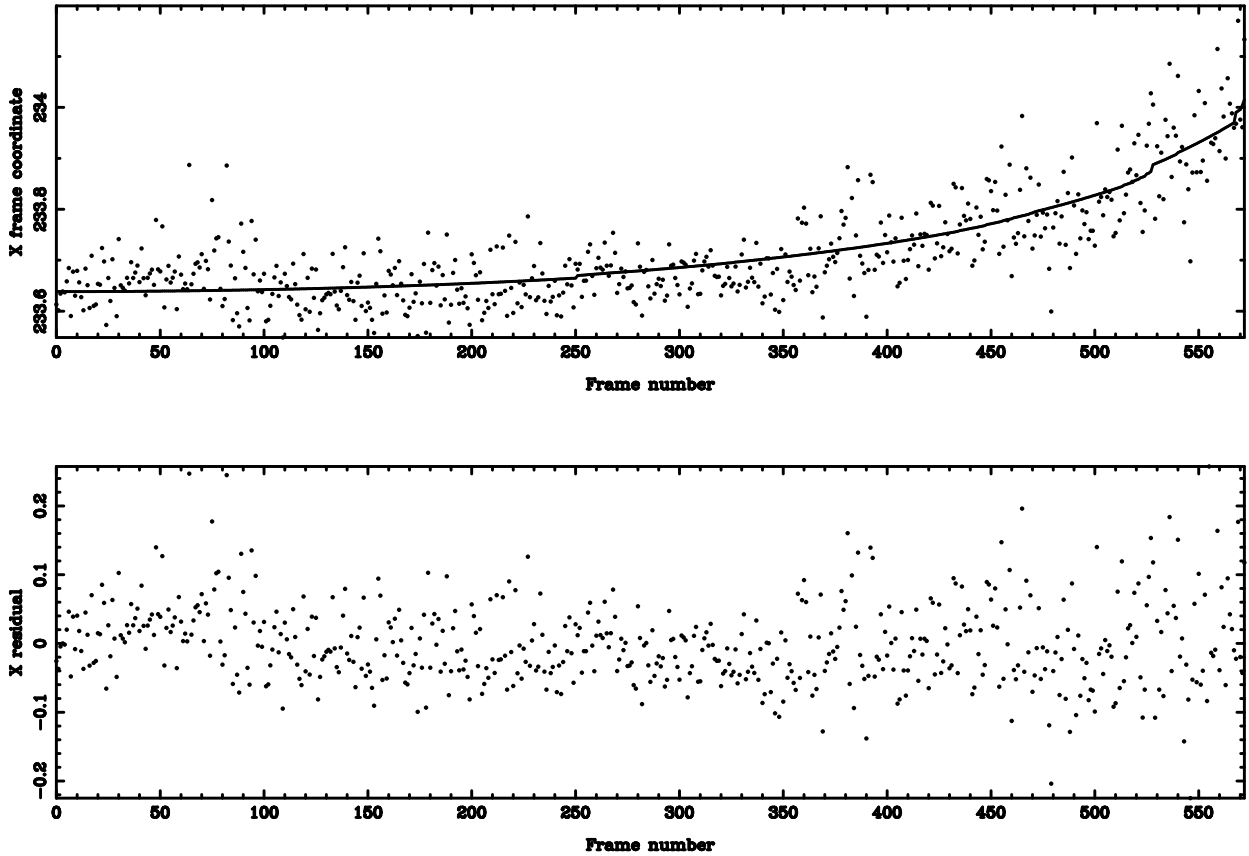}

\caption{The measured pixel position of the `notch' as a function of
Frame Number with the solid line showing the linear fit with airmass
(upper figure) and the residual between the data and the fitted notch
position as a function of Frame Number (lower figure).
\label{notchfit}}

\end{figure}

The wavelength zero point is determined from the position of atmospheric
absorption due to the Fraunhofer A band of Oxygen (hereafter referred to
as the 'notch') at 7620 Angstroms. Measuring the notch pixel position
with respect to the zeroth order image of the star is performed by
fitting a gaussian to the notch in the extracted spectrum.  The position
of the notch is expected to smoothly change with airmass during the
night, and so we take the fitted value of the notch position to be our
wavelength zero point.  Figure \ref{notchfit} shows the centroid of the
notch as a function of frame number and the residuals after taking off a
linear fit in effective airmass $a$.  The RMS between the centroid fit
and measured centroid of the notch is $6.0\times10^{-2}$ pixels.
Assuming that the measured notch centroids are independent of each
other, the mean centroid error is reduced by the square root of the
number of frames used in the notch fit. This gives a mean notch centroid
error of $6.0\times 10^{-2} / \surd{573} = 2.5\times 10^{-4}$ pixels (
$3.5\times 10^{-2}$\AA).

As the star moves from the zenith an extra component of dispersion due
to colour-dependent refraction in the the Earth's atmosphere is added to
the dispersion of the grating and the wavelength solution changes
smoothly and continuously throughout the night. This atmospheric
dispersion adds vectorially to the transmission grating dispersion to
produce the observed wavelength solution on a given frame.

At the zenith, the grating is assumed to have a low-order polynomial
wavelength solution where distance from the zeroth order image is
monotonic with wavelength.  A planetary nebula (PNe) of small angular
extent (IC 2165) is observed and provides wavelength calibration for the
extracted spectrum in the form of emission lines of $\rm H\alpha $ and
O[\rm{III}]. This observation was taken at a high airmass, and so the
line positions are corrected using an atmospheric refraction model back
to the zenith. We use a model of atmospheric refracton for the Earth's
atmosphere from \cite{fil82}, taking estimated values for the
temperature and pressure from the observatory log sheets. The PNe
wavelength data is corrected back to the zenith, and the resultant
dispersion solution is linear within measuring errors, implying that the
grating dispersion is linear with wavelength. 

Using the wavelength solution for the grating and the vector correction
for dispersion in the atmosphere at the required airmass, the spectra
are resampled using cubic spline interpolation into $50$\AA~wavelength
bins from $4000-8000$\AA. This spectral range is limited by the
efficiency of the detector in the blue end of the spectrum, and the
second order spectrum appearing beyond $8000$\AA.

\section{The Atmospheric Model}

For a single observed stellar spectrum $I(a,\lambda)$ as observed from the
ground at effective airmass $a$, we assume an atmospheric model of the form:

\begin{equation}
I(a,\lambda)=I(\lambda).e^{-a.\tau(\lambda)}
\end{equation}

where $I(\lambda)$ is the stellar spectrum for an effective airmass of
zero, i.e. above the atmosphere, and $\tau(\lambda)$ is the atmospheric
absorption.

For an eclisping binary at secondary eclipse, the unobscured fraction of
the secondary's disk varies from 1 at first contact to 0 at second
contact, remaining at 0 until it emerges from behind the primary at
third contact and rising up to 1 at fourth contact. This function is
represented as $p(t)$. For the geometry of the HD~209458 system, the
reflected light also varies as the phase of the planet varies, but the
$p(t)$ is a good approximation to the actual phase for our data set.

The usual definition of $a=\sec z$ holds true for a semi-infinite
plane-parallel atomsphere. We keep the semi-infinite atmosphere and take
account of the curvature of the Earth. A simple function relates the
zenith distance $z$ of the star to an effective airmass $a$:

\begin{equation}
a_{eff} = \frac{R}{r} . \left( 
\sqrt{\cos^2 z + 2.\left( \frac{r}{R}\right) + \left(\frac{r}{R}\right)^2}
 - \cos z \right)
\end{equation}
 
Where the radius of the Earth is $R=6378\rm{km}$ and the height of a
homogeneous atmoshpere (finite height, constant temperature equal to the
surface temperature and constant density equal to surface density) at 15
Celsius is $r=8.43\rm{km}$ \citep{cox99}. Using equation 2 we calculate
the effective airmass as a function of time. 

Given knowledge of the ingress and egress times for an eclipsing binary
system and the geometry of the system, a model for the secondary eclipse
as observed on the ground is constructed:

\begin{equation}
I(a,t, \lambda) = I(\lambda).[1 +
f(\lambda).p(t)]e^{-a(t).\tau(\lambda)}
\end{equation}

Here we define $f(\lambda)$, the ratio of the secondary's spectrum to
that of the primary star. For reflected light from an ESP, this is
proportional to the planet's albedo.

Taking the logarithm of this equation and realising that $\ln(1+x)\simeq
x$ for $x \ll 1$ gives:

$$\ln I(a,t,\lambda) = \ln I(\lambda) + f(\lambda).p(t) -
a(t).\tau(\lambda)$$

This linear equation is then fitted numerically with linear methods to
produce estimates of $I, f, \tau$.  Errors on each of the free parameters
are estimated by simulating data for the star system assuming
limitations due to Poisson distributed noise. For a given set of $I_0$,
$\tau_0$, and $f_0$ at a particular wavelength, an ideal noiseless $I(a)$
is constructed and noise commensurate with our observations is added.
The noise is estimated by taking the r.m.s. of the difference of the
fitted light curve and the actual data. A linear fitting routine then
extracts ``measured'' values of $I,\tau,f$. Repeating this process with a
different sample of Poisson noise produces distributions of $I_m\pm
\delta I, \tau_m\pm \delta \tau, f_m\pm \delta f$, where the $\delta-$
quantities correspond to 1-sigma limits of the resultant Poisson
distributions, and these are the error bars on our results.

\begin{figure}
\includegraphics[angle=270,width=\columnwidth]{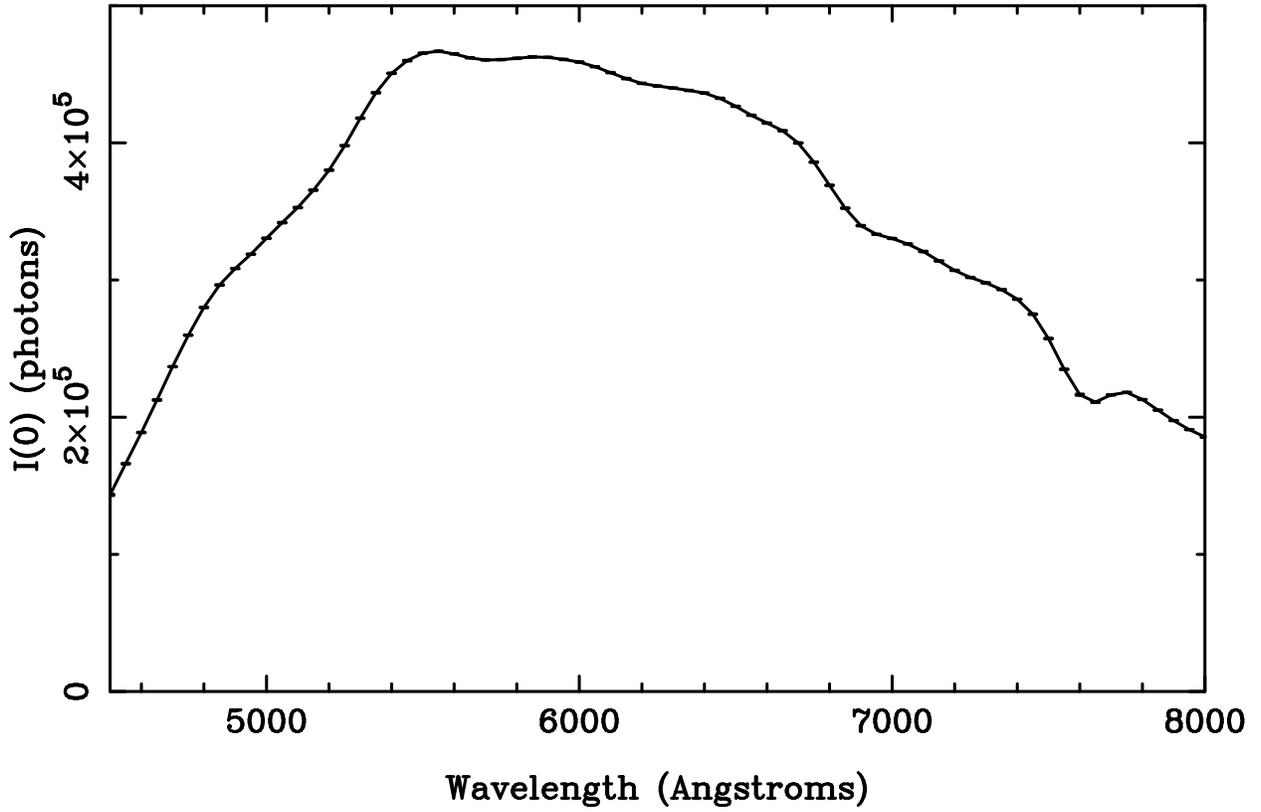}

\caption{The extracted spectrum of HD~209458 for zero airmass. Error
bars shown on the figure.  \label{specI0}}

\end{figure}

\begin{figure}
\includegraphics[angle=270,width=\columnwidth]{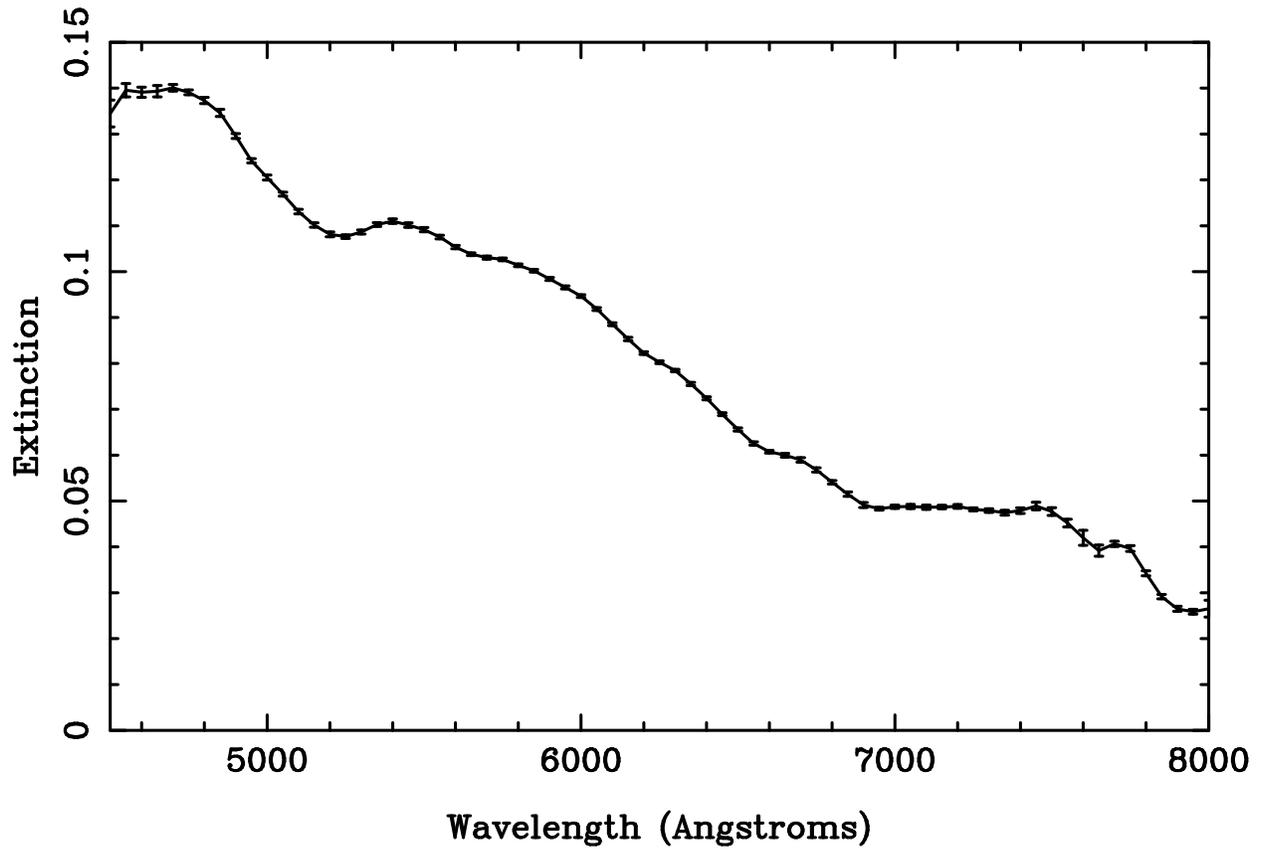}

\caption{The extracted spectrum of the atmospheric absorption using
HD~209458.  \label{spectau}}

\end{figure}

\begin{figure}
\includegraphics[angle=270,width=\columnwidth]{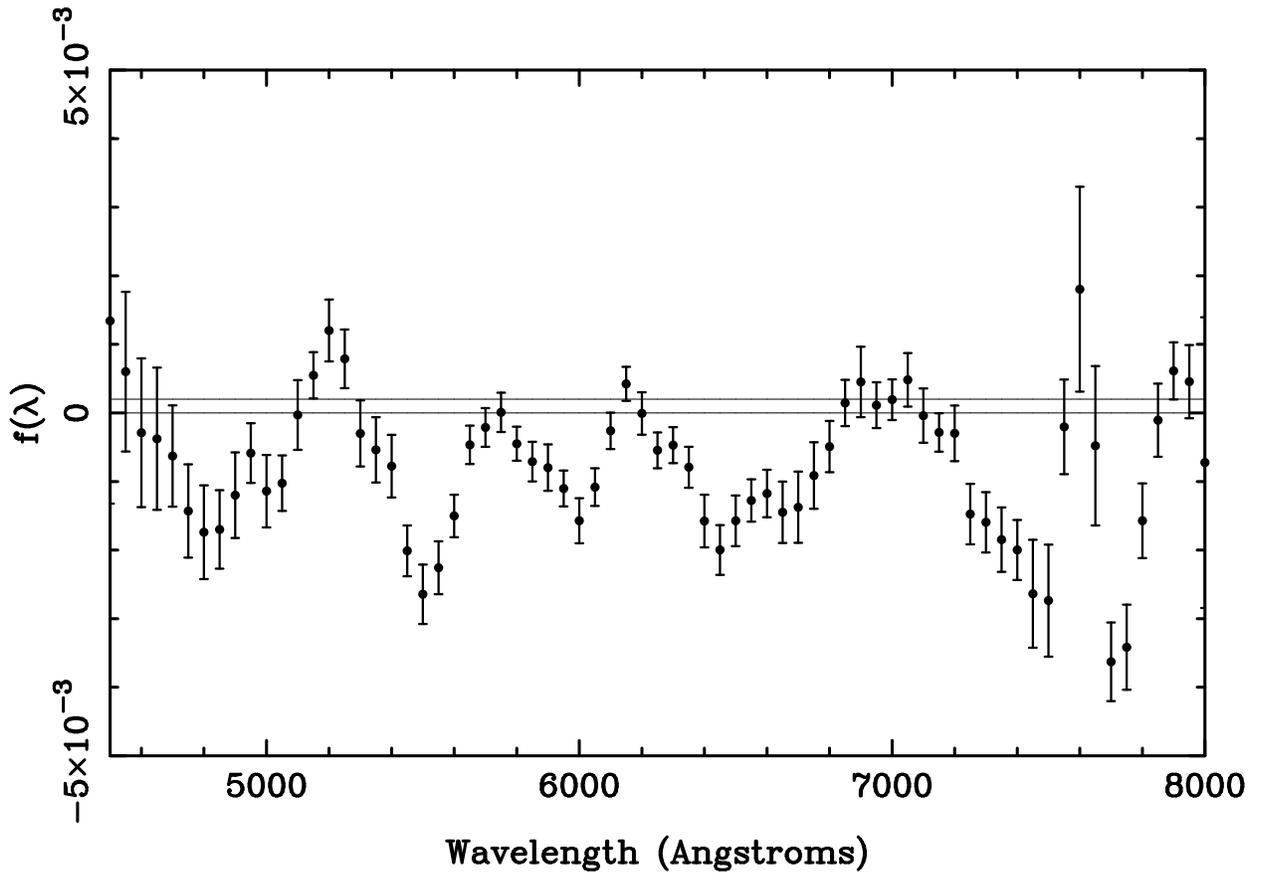}

\caption{The extracted reflected light spectrum of the planet HD~209458b. 
\label{specp}}

\end{figure}

Figure \ref{specI0} and \ref{spectau} show the fitted spectrum of
HD~209458 and the derived extinction curve of the atmosphere. The
derived spectrum for the planet HD~209458b is shown in Figure
\ref{specp}.  The mean planet flux over the passband 5000-7000\AA~is
$-7.0\pm8.4 \times 10^{-4}$, the sensitivity of which is approximately 4
times larger than the maximum reflected light fraction of $2\times
10^{-4}$.

\begin{figure}
\includegraphics[angle=270,width=\columnwidth]{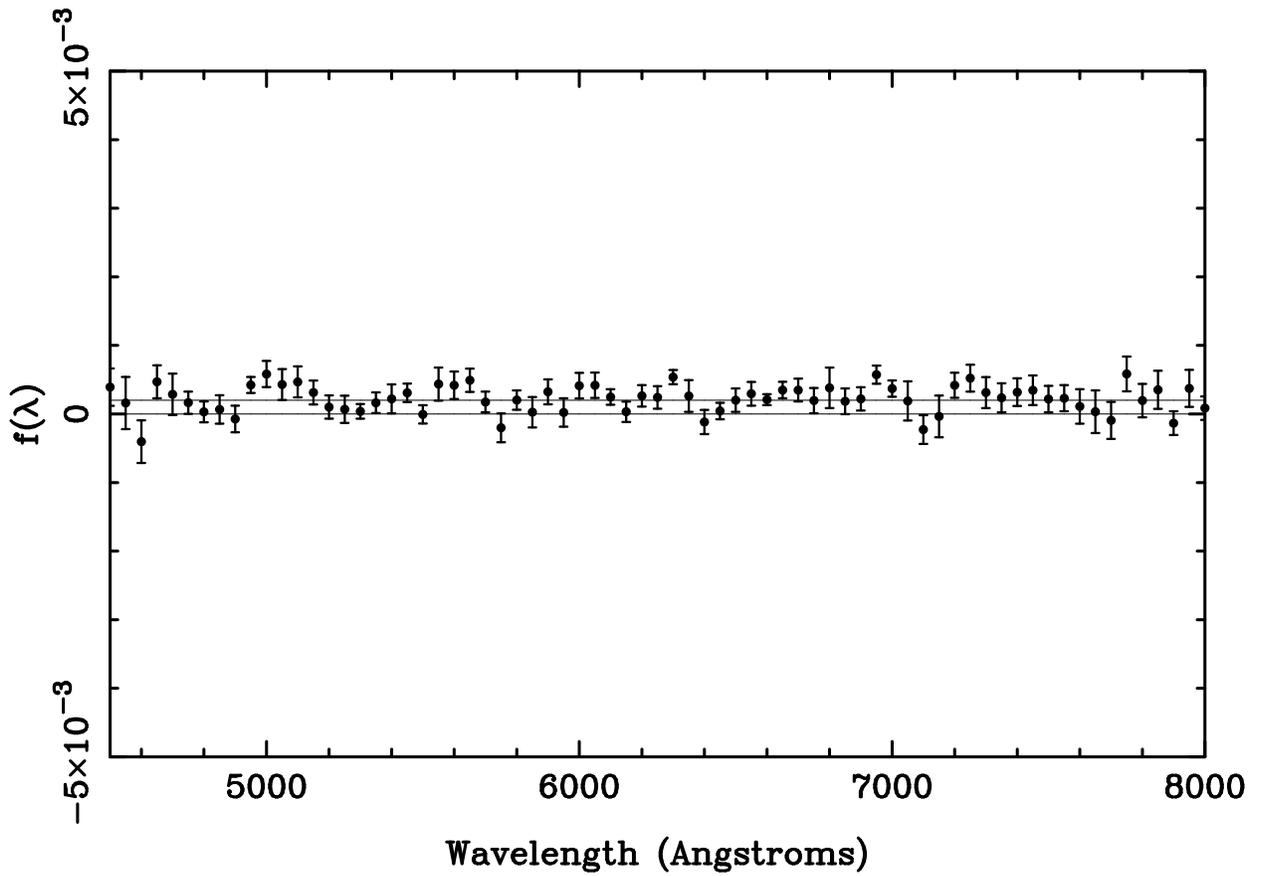}

\caption{Simulated extracted planet spectrum for an ideal atmosphere
with photon noise. The upper and lower lines represent
$f(\lambda)=2\times 10^{-4}$ and $f(\lambda)=0$ respectively.
\label{simphoton}}

\end{figure}

Using our model of the atmosphere we construct a set of model spectra
for the measured range of airmass $a(t)$ and planet phase $p(t)$. For
$I(\lambda)$ and $\tau(\lambda)$ we use the values derived in the
extraction of HD~209458 and we assume a constant reflected light
fraction $f(\lambda)=2\times 10^{-4}$.  Adding only photon shot noise
results in an extracted spectrum as shown in Figure \ref{simphoton}. The
measured planet flux is detected with a level of
$(2.0\pm1.8)\times10^{-4}$ per 50\AA~bin. Averaged over the 40 bins in
the range of 5000-7000\AA~this results in an accuracy of $2.8\times
10^{-5} $ in the limit of photon noise.

\begin{figure}
\includegraphics[angle=270,width=\columnwidth]{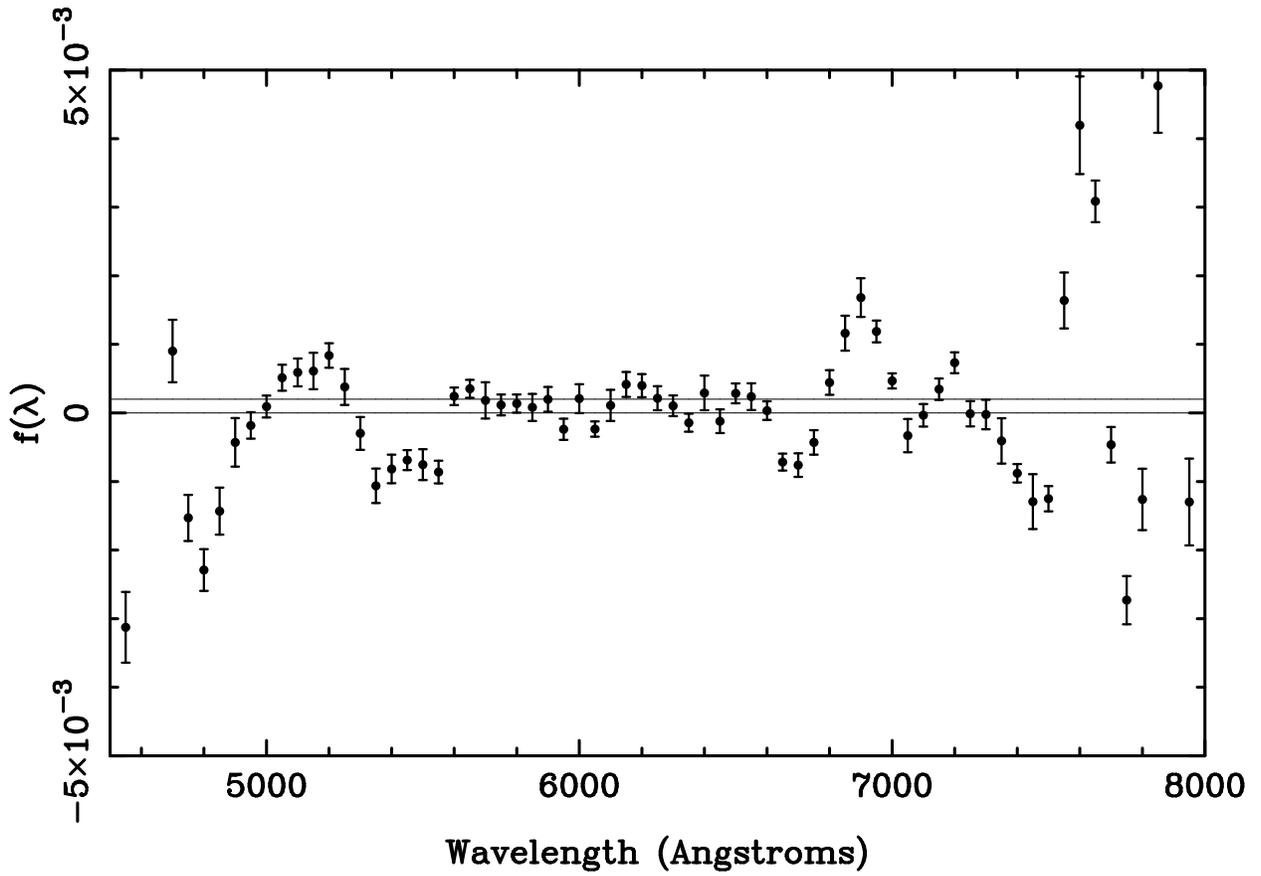}

\caption{Simulated extracted planet spectrum for ideal atmosphere with
photon noise and atmospheric blurring.  \label{simphotblur}}

\end{figure}

\begin{figure}
\includegraphics[angle=270,width=\columnwidth]{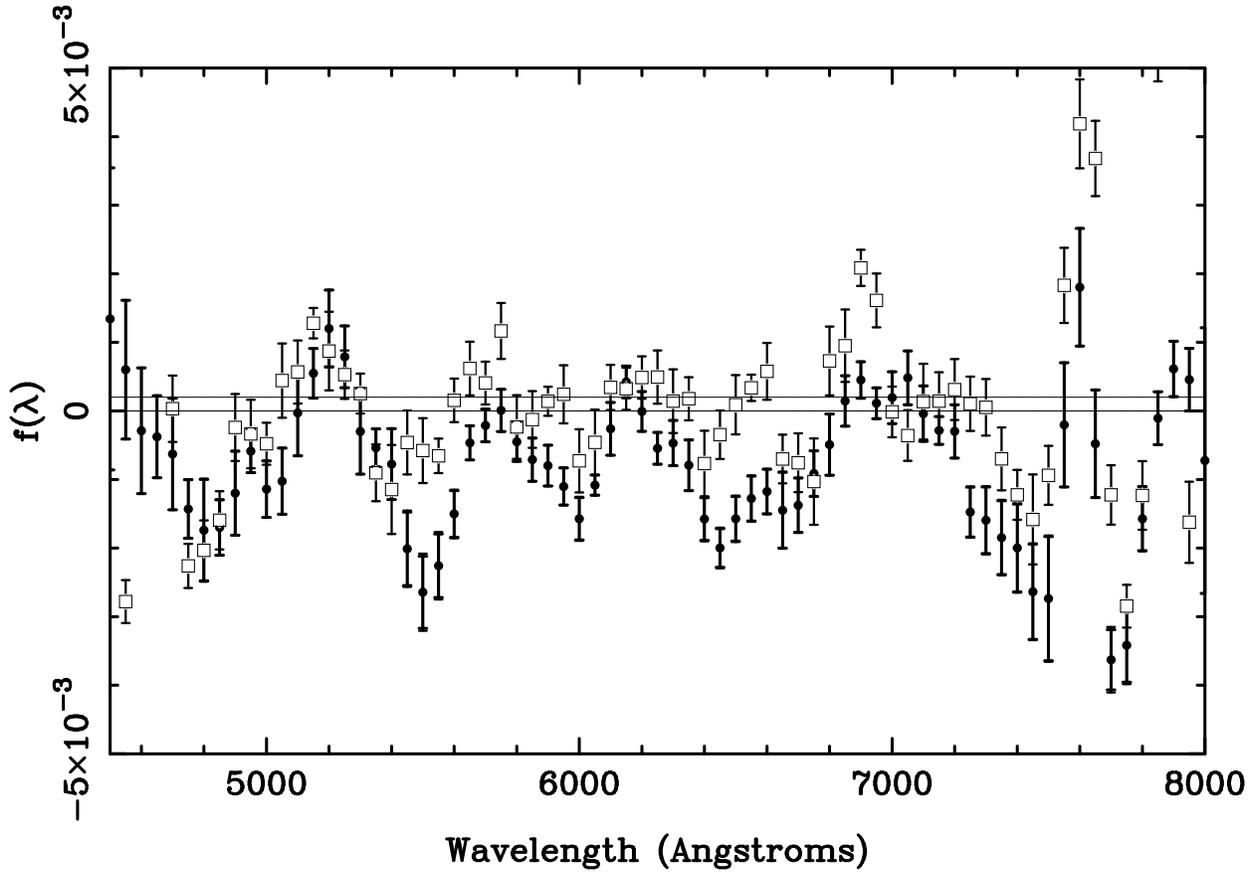}

\caption{Simulated and actual extracted model planet spectrum for ideal
atmosphere with photon noise, scintillation noise and atmospheric
blurring. Circles are the actual data, and white squares are the
simulated data. The dominant effect appears to be from variable
atmospheric blurring of the spectrum.\label{simscint2}}

\end{figure}

Making the assumption that the PSF along the spectra is the same as the
PSF measured across the spectrum, we apply atmospheric seeing effects to
the data set. The extracted planetary spectrum (Figure
\ref{simphotblur}) shows a significant systematic error similiar in sign
and magnitude to that seen in our measured data. Our exposure times are
short enough that we are also affected by scintillation noise and
including this contribution produces the spectrum shown in Figure
\ref{simscint2}. 

A simulated spectrum can be created with no planetary signal which has
the same systematic errors as the actual data, as shown in Figure
\ref{simscint2}. This can be used to normalise the signal detected in the
data.  We create a normalised spectrum $f_{norm}(\lambda)$ with:

$$f_{norm}(\lambda) = \frac{1+f_{meas}}{1+f_{model}} $$

where $f_{meas}$ is our observed planet spectrum, and $f_{model}$ is the
simulated planet spectrum. 

\begin{figure}
\includegraphics[angle=270,width=\columnwidth]{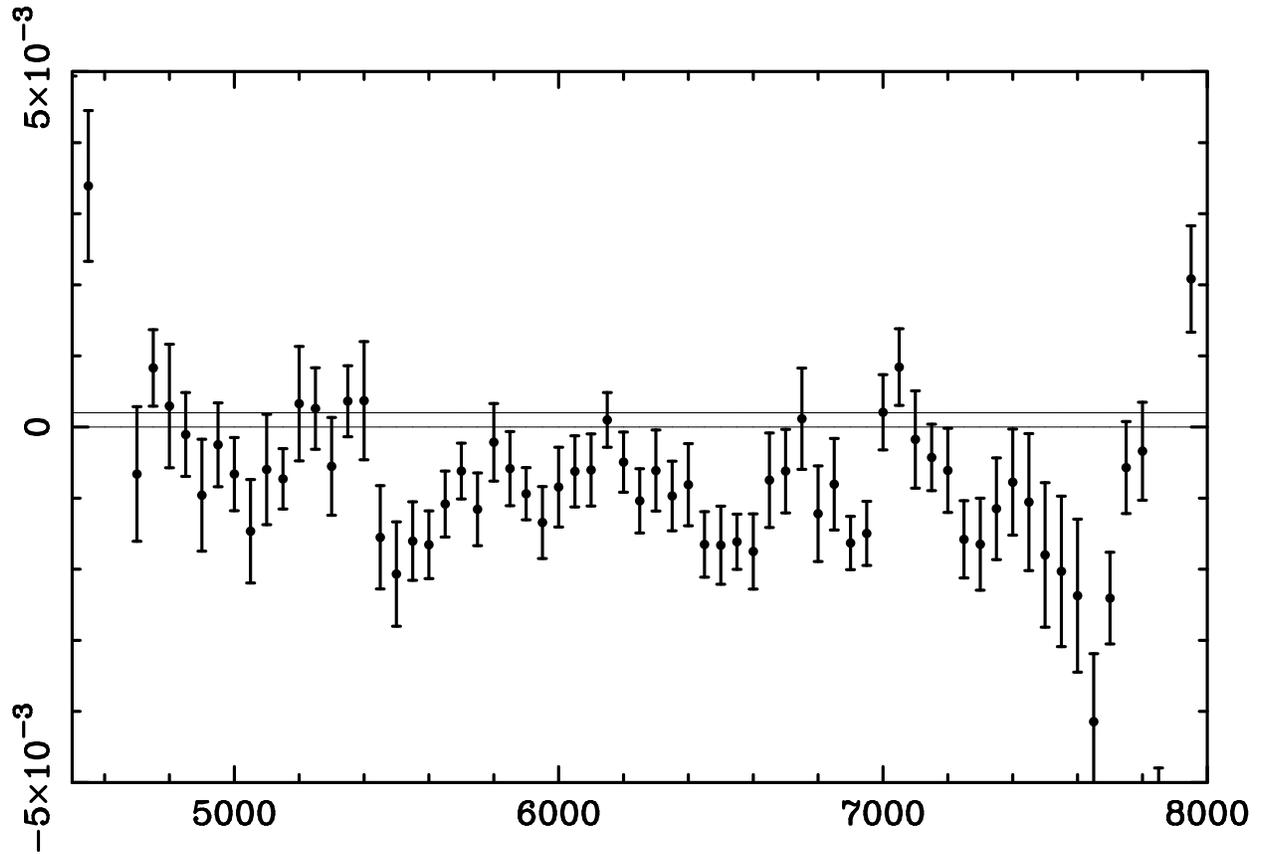}

\caption{Ratio of the real signal to the artificial signal with photon
noise, scintillation noise and atmospheric blurring. Compare with Figure
\ref{specp} for the reduction of the systematic noise.  \label{divide}}

\end{figure}

The resultant spectrum in Figure \ref{divide} shows a reduction in
systematic effects and gives a mean fractional planet flux of
$-8.2\pm6.2\times10^{-4}$ averaged over the bandpass of 5000-7000\AA.

\section{Limits to the Precision}

The fundamental limit to the precision of the photometric measurement in
each wavelength bin is set by the number of photons collected.  The
uncertainty in the intensity is governed by Poisson statistics, and is
thus given by the square root of the photons collected per wavelength
bin per exposure. As can be seen in Figure \ref{specI0} the peak flux
per exposure per 50\AA~wavelength bin at an airmass of zero is
approximately $4.0\times 10^{5}$ photons for most of the
5000-7000\AA~portion of the spectrum. Taking into account the absorption
of the atmosphere averaged over the night, this figure lowers to
$3.5\times 10^5$ photons, creating an uncertainty of $1.69\times10^{-3}$
in the intensity.  However, no correlation exists between wavelength
bins or exposures, allowing this uncertainty to be reduced.  For the 240
exposures taken while the planet was eclipsed by the star, and the 200
taken when the planet was not, these give uncertainties of
$1.12\times10^{-4}$ and $1.09\times10^{-4}$ which when combined in
quadrature give a photon-limted uncertainty per wavelength bin of
$1.56\times10^{-4}$, comparable to the $1.8\times10^{-4}$ from the
photon-limited model in the previous section.

The measured variations in intensity per frame are also expected to be
affected by scintillation of the starlight. \cite{you67} derived the
variations in intensity due to scintillation.  The uncertainty in
intensity for typical scintillation is given by the equation

$$\sigma = S_0.D^{-2/3}(\sec Z)^{1.75}e^{-h/h_0}/(t)^{1/2} $$

where $D$ is the telescope aperture in meters, $t$ is the exposure time
in seconds, $Z$ is the zenith angle, $h$ is the height of the
observatory (2.5 km for the Kuiper telescope), and $h_0$ is the
turbulence weighted atmospheric altitude (typically 8 km). $S_0$ is the
normalization constant for the effect, which is typically 0.003, but can
vary due to the strength and height of the turbulence.  Work by
\cite{dra98} have verified this equation for various timescales and
aperture sizes, showing that the strength of scintillation varies (as
parameterized by $S_0$ in the equation above) and is not well correlated
with the local seeing. For the 5 second frames, at airmass 1.5 each
frame would have a variation of 1.5$\times10^{-3}$.  As indicated by the
square root dependence on exposure time, successive images are not
correlated.  Thus for the 240 exposures taken while the planet was
eclipsed by the star we expect the scintillation-induced uncertainty per
wavelength bin to be 1.3$\times10^{-4}$.  However, studies by
\cite{dra97} and \cite{rya98} do show that the the effects of
scintillation can be considered achromatic in their effect on a
spectrum. That is, the variations in intensity are correlated between
wavelength bins.

\begin{figure}
\includegraphics[angle=270,width=\columnwidth]{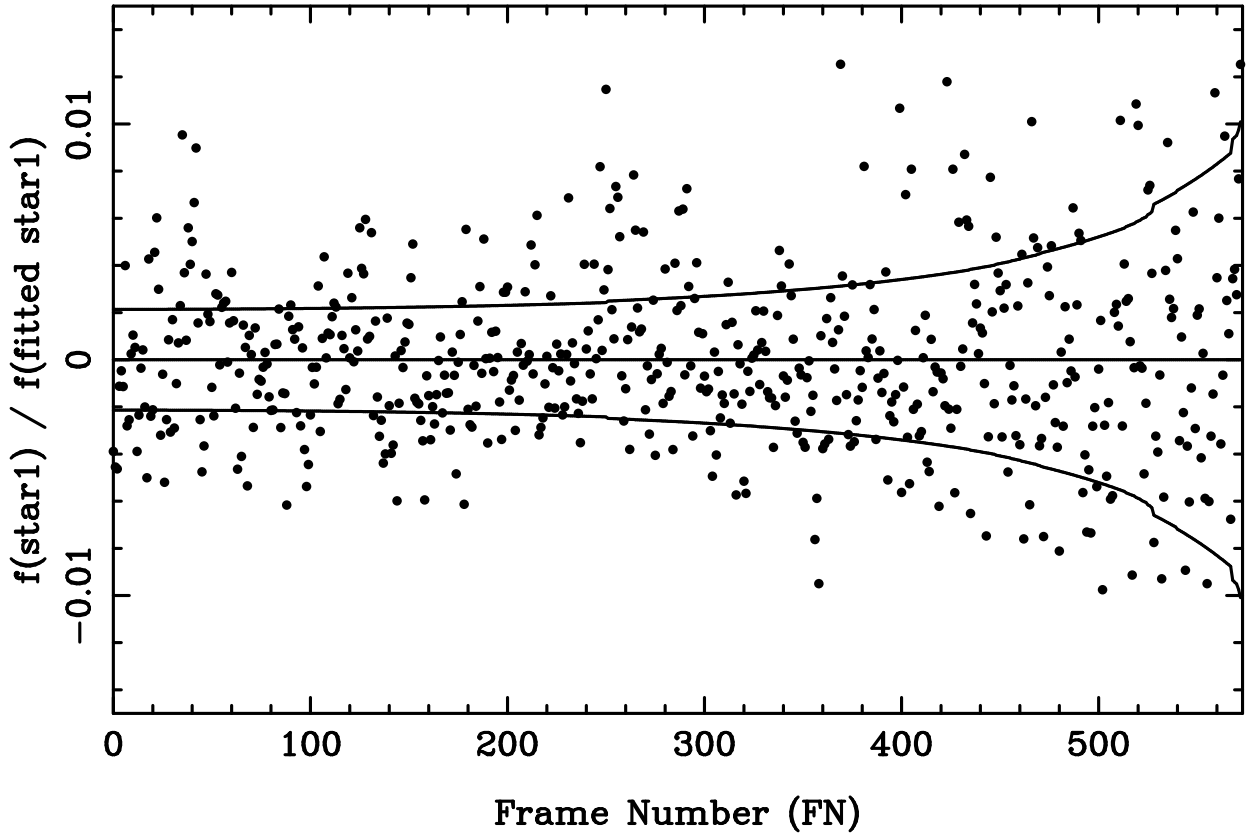}

\caption{Star 1 flux fitted with a low-order function to isolate the
short-term variations and so estimate the scintillation noise. The upper
and lower solid curves represent 1-$\sigma$ best fit estimates of
scintillation and photon shot noise.  \label{star1noise}}

\end{figure}

The value of $S_0$ changes significantly between nights and observing
sites \citep{dra97}, and so we use the measured flux of Star 1 to
determine the value of this scintillation constant for our data. We
calculate the flux of Star 1 as a fuction of airmass and fit an
exponential curve to the data, assuming a single value for $\tau$. From
our observed data, we know that $\tau$ changes as a function of
wavelength, and the difference of the fit and the data show this as a
low-order residual with a peak of approximately 0.5\% of star 1's flux
at 1 airmass. A low-order curve is fitted to remove this remaining flux,
and for each observed value of the star 1's flux, an estimate of the
poisson noise and the scintillation noise is made (see Figure
\ref{star1noise}). The scintillation coefficient that best fits the
scatter in star 1's flux is $0.006\pm 0.002$, and the short-term
variations in the flux are attributed to variations in $S_0$ during the
night.  This corresponds to a scintillation-induced uncertainty of
approximately $3\times10^{-3}$ per frame or 2.6$\times10^{-4}$ over
the transit of the planet.

As can be seen by the comparison of photon and scintillation noise, for
nominal scintillation the spectral resolution is appropriate for nearly
equal noise contributions from each. Higher spectral resolution would
cause photon noise to dominate for each wavelength bin, while lower
resolution would cause scintillation noise to dominate. Thus a spectral
resolution of 30-60 is suitable for balancing these two sources of
error. The photon noise scales inversely with the diameter of the
telescope while the scintillation scales inversely as the 2/3 power of
the diameter. In general this means that the optimum resolution for
balancing these sources of noise (for HD~209458) is given by
$R_{opt}\sim 30\times D^{1/3} $ where D is the telescope diameter in
meters. This suggests that even for a large aperture telescope (8-10 m)
the optimum resolution is relatively low (around 100) even for planets
around fairly bright stars.

In addition to random variations in intensity we expect systematic
effects which may cause variations in intensity from several sources.
For the low-resolution spectrum, each wavelength bin is actually a blend
of a range of wavelengths corresponding to approximately 150\AA.
This blending of wavelengths creates a systematic curvature to the
extinction fit for regions of the spectrum where either the atmospheric
extinction is changing or the intensity is changing. 

From comparison with our simulated data it is clear that the dominant
effect limiting our precision is the effect of wavelength blending in
our spectra, and the inability to fit this properly. As shown in Figure
\ref{divide} this limits measurements of the planet flux in the
5000-7000\AA~range to a level of approximately 0.1\%.

The discrepancy between the model fit and the extracted spectra show
that the PSF blurring along the spectral dispersion axis may not be well
represented by the transverse profile.  For a comprehensive image
deconvolution method to be used on the data frames, careful calibration
of the PSF as a function of position on the detector, and relative
position of star and grating and detector must be built up for a range
of seeing conditions and all with a signal to noise of at least the
required sensitivity. Even so, we have looked at various deconvolution
techniques based on minimising a function that trades chi squared of a
trial fit with the trial spectrum's entropy, as used in many maximum
entropy codes. Our success, however, has been severely limited by the
lack of knowledge of the PSF introduced by the atmosphere and grating.

\section{Conclusions}

The current approach to obtaining data of HD~209458 results in very
inefficient observing. For every 5 second exposure, approximately 30
seconds of real time is needed in order to read out the CCD, resulting
in 16\% efficiency to the observing. The exposures are limited to 5
seconds currently to avoid saturating the CCD.  However, an increase in
the efficiency can be made by spreading the light out perpendicular to
the direction of dispersion with a cylindrical lens.  This will allow
much longer exposures resulting in an efficiency $>50\%$. This would
result in a decrease in the noise associated with both the photon and
scintillation noise allowing reduction of these quantitie to well below
the expected signal due to a planetary secondary eclipse.

The power of the technique we describe is that it requires very little
in extra calibration data - an estimate of the atmospheric blurring is
made from transverse cuts across the spectrum, and each frame thus
provides its own calibration data. 

Precision spectrophotometry from ground-based telescopes is an
attractive technique for probing secondary eclipses of planetary orbits,
but the current work shows the technique is limited by systematic
effects of atmospheric extinction to precision of 0.1\%. It is possible
that future refinements of the technique will allow an improvement by
calibrating out the effect of the atmosphere through appropriate
standard star observations.  This will allow explorations of intensity
variations down to the level set by the photon and scintillation noise
of the star. 

\acknowledgments

We thank Bob Peterson and the Mount Bigelow crew for their patience with
our constant requests for observing time and modifications to the
telescope.

\clearpage

\clearpage

\clearpage

\clearpage

\clearpage

\clearpage

\clearpage

\clearpage

\clearpage

\clearpage

\clearpage

\clearpage

\clearpage

\clearpage

\clearpage

\clearpage

\clearpage

\clearpage

\begin{deluxetable}{lcc}
\tabletypesize{\scriptsize}
\tablecaption{Ephemeris for HD~209458b anti-transit on 12 Nov 2001.
\label{ephem_table}}
\tablewidth{0pt}
\tablehead{ \colhead{} & \colhead{Universal Time} & \colhead{Airmass} }
\startdata
Start of ingress& 02:17:33 UT & 1.03 \\
End   of ingress& 02:41:53 UT & 1.04 \\
mid transit     & 03:46:50 UT & 1.64 \\ 
Start of  egress& 04:51:46 UT & 1.87 \\
End   of  egress& 05:16:07 UT & 2.06 \\
\enddata

\end{deluxetable}

\clearpage

\begin{deluxetable}{lcc}
\tabletypesize{\scriptsize}
\tablecaption{Noise Statistics of the flat fields.
\label{flats_table}}
\tablewidth{0pt}
\tablehead{ \colhead{} & \colhead{Twilight sky} & \colhead{Halogen lamp} }
\startdata
Number of individual frames         & 18                & 114 \\
Mean counts per individual frame    & 35632             & 51718 \\
RMS of  frames                      & $117.9\pm21.5$    & $135.8\pm8.7$ \\
SNR for frames                      & $3.31\pm0.60\times10^{-3}$    & $2.62\pm0.17\times10^{-3}$ \\
SNR for combined frames             & $7.8\pm1.4\times10^{-4}$    & $2.45\pm0.16\times10^{-4}$ \\
\enddata

\end{deluxetable}

\clearpage

\begin{deluxetable}{lcc}
\tabletypesize{\scriptsize}
\tablecaption{Throughput estimates\label{thput_table}}
\tablewidth{0pt}
\tablehead{ \colhead{Optic} & \colhead{Estimated transmission at 6000\AA} }
\startdata
Atmosphere & 0.90\tablenotemark{a} \\
Telescope & 0.85 (estimated)\\
Dewar windows & 0.92\tablenotemark{b}\\
Grating (1st order) & 0.32\tablenotemark{c}\\
CCD Q.E. & 0.40\tablenotemark{d} \\
Total efficiency & 0.090 (estimated) \\
\enddata

\tablenotetext{a}{takes value of $\tau=0.10$ from atmosphere model}
\tablenotetext{b}{assumes 4 glass/air AR coated surfaces}
\tablenotetext{c}{from Roy and Milton grating catalogue}
\tablenotetext{d}{from Univerity of Arizona CCD Laboratory data}

\end{deluxetable}

\end{document}